\documentclass[]{raa}            % referee version: for submission
\usepackage{graphicx,times}
\usepackage{natbib}

\providecommand{\bm}[1]{\mbox{\boldmath$#1$\unboldmath}}

\usepackage{amssymb}
\usepackage{amsfonts}
\usepackage{txfonts}
\usepackage{natbib}

\bibpunct{(}{)}{;}{a}{}{,}
\begin{document}

\title{Stellar kinematics and populations out to 1.5 effective radius in 
the elliptical galaxy NGC4636}
 \volnopage{ {\bf 2011} Vol.\ {\bf 000} No. {\bf XX}, 000--000}
   \setcounter{page}{1}
\author{S. B. Pu\inst{1,2,3}
 \thanks{e-mail:pushibi@hotmail.com}
  \and   Z. Han\inst{1,3}
         }
 \institute{National Astronomical Observatory/Yunnan Observatory, 
the Chinese Academy of Sciences, Kunming 650011, China
 \and Graduate University of Chinese Academy of Sciences, Beijing 100049, China
 \and Key Laboratory of the Structure and Evolution of Celestial Objects,
Chinese Academy of Sciences, Kunming 650011, China
 }
%********************************************************************
\abstract{We present high quality long slit spectra along the major and minor 
axes out to 1.5 effective radius ($R_e$) of the massive galaxy NGC4636 taken by 
Hobby-Eberly Telescope (HET). Using Fourier Correlation Quotient (FCQ) method,
we measured the stellar line-of-sight velocity distribution along the axes. 
Furthermore, six Lick/IDS indices ($H\beta,Mgb,Fe_{5015},Fe_{5270},Fe_{5335},Fe_{5406}$) 
are derived from the clean spectrum. By comparing the measured absorption line strengths with the 
predictions of Simple Stellar Populations (SSP) models, we derived ages, 
total metallicity and 
$\alpha$ abundance profiles of the galaxy. This galaxy presents old and 
$[\alpha/Fe]$ over abundant stellar populations.
Indeed, using the SSP model, we obtained the broadband color profiles. The 
theoretical colors match well with the measured colors and present red sharp 
peaks at the galaxy center. The sharp peaks of the colors are mainly shaped by the high 
metallicity in the galaxy center. Interestingly, the galaxy has steep negative
metallicity gradients, but trend flattens outwards. This result likly suggests  
that the center and outer regions of the galaxy formed through different formation process.   
}
\maketitle
\section{Introduction}
\label{sec_intro} 
The formation and evolution of massive, early-type galaxies constitutes a 
long-standing and crucial problem in cosmology \citep{sanchez07}.
According to the classic \emph{monolithic-collapse} model for the formation 
and evolution of early-type galaxies \citep{tinsley72,larson75,tantalo96}, 
the early-type galaxies formed 
most of their stars during a single short and highly efficient star formation 
event in the early universe. 
This model is strongly supported by the extremely small scatter of 
the observed color magnitude relation of elliptical galaxies. This uniformity 
of stellar populations in ellipticals is also supported by the 
Fundamental Plane \citep{dressler,djorgovski,bender92a,saglia93}. 
Moreover, the tightness of the \begin{math}Mg-\sigma\end
{math} relation for massive elliptical galaxies observed in the local universe 
\citep{bender93,sanchez07} but holding up to the 
intermediate redshift \begin{math}z\thickapprox 1\end{math} 
\citep{ziegler,bender98} requires a picture of a short, 
and highly efficient star formation process at high redshift and passive 
evolution since then.

In contrast, according to the \emph{hierarchical merging} scenario 
\citep{white78,kuaffmann}, the massive early-type 
galaxies are expected to have formed through multiple mergers and the 
accretion of smaller objects over an extended period 
\citep{white91,somerville,delucia}. 
This formation scenario has been observationally confirmed; COMBO-17 and DEEP2 
surveys show that the number density of red galaxies has increased since 
redshift z=1 \citep{bell04,faber07}.  
Furthermore, giant galaxies show boxy isophotes and anisotropic dynamics, and 
more massive galaxies are more radio-loud,
more strong X-ray emitters, more frequently disturbed. On the contrary,
normal and low luminosity elliptical galaxies rotate rapidly, are nearly 
isotropic, show disky distorted isophotes and cuspy inner profiles. 
The properties of the former can be explaned in dissipationaless mergers, 
while the latter are recovered successfully with dissipational mergers
\citep{nieto,bender89,bender92b,barnes92,mehlert}. Recently,
\citet{kuntschner,thomas05,collobert,bernardi06,clemens,rogers}
showed that early-type galaxies in low density and in high density environments
might exhibit different formation ages, and
\citet{lisker,sanchez09,matkovic} 
found evidence that lower mass galaxies have more extended star formation histories.  

Radial profiles of the kinematics, colors, ages and metallicities of the 
stellar populations are the efficient tools to study galaxy formation 
scenarios. In standard closed-box models of chemical enrichment, the 
metallicity is a function of the yield and of how much gas has been locked 
after star formation has ceased \citep{tinsley80}.  Therefore, the
metallicity strongly depends on the dynamical parameters. For instance, 
for galaxies formed via a \emph{monolithic-collapse}, stars formed in all regions 
during the collapse and remain in their orbits with little movement inward, 
whereas the gas dissipates into the center, being continuously enriched by the 
evolving stars. Therefore, stars formed in the center are predicted to be more 
metal rich than those in the outer regions. So far the galaxies should have 
steep radial metallicity gradients 
\citep{larson76,thomas99,sanchez07}. 
On the other hand, major mergers, which come along with 
\emph{hierarchical merging}, will dilute stellar population 
gradients \citep{white80,kobayashi,hopkin,tortora}. Therefore, more flat population gradients are expected 
within this picture. Accordingly, the observation of stellar population 
gradients and their connection to dynamical parameters can give crucial 
insight into the formation paths of individual galaxies.

This work is aiming to deeply investigate the stellar kinematics and populations 
of the massive galaxy \object{NGC4636}. \object{NGC4636} is an E/S0 galaxy, 
located at about 2.8Mpc southeast from the Virgo center and 14.7 Mpc 
[$(m-M)_0$ = 30.83 $\pm 0.13$] from us. Effective radius, ellipticity, and 
position angle of major axis of \object{NGC4636} are 
$R_{e} = 88.5\arcsec$, $\epsilon_{e}$ =0.256, and PA = $150^{\circ}$, respectively 
\citep{tonry,rampazzo,kim,schuberth}. \object{NGC4636} is considered to be a major member 
of a small group falling into the Virgo center \citep{nolthenius}. 
Although, it is relatively less luminous ($M_v$ = -21.7 mag, ) among the 
gEs in Virgo, but \object{NGC4636} shows several interesting features. For instance, 
\object{NGC4636} is found to be very bright in X-rays 
($L_X$ $\sim$ $10^{41}$ ergs $s^{-1}$), with unusual feature in the hot interstellar 
medium (ISM) \citep{osullivan,kim,posson}.
The galaxy \object{NGC4636}, has boxy isophotes \citep{rembold} 
and does not show rotation both along major and minor axes 
\citep{davies,bender94,rampazzo}. Indeed, 
\object{NGC4636} is one of the best targets for studying kinematics of globular 
clusters since it has an anomalously large number of GCs 
\citep{dirsch,chakrabarty,park,lee10}.

There is a number of works which focus on the study of
kinematic profiles, line strength indices and stellar population
parameters in \object{NGC4636}
\citep{davies,bender94,tantalo98,rampazzo,annibali06,li}.  However, 
the previous measurements are concentrated within $R_{e}$/2 of galaxies.  
From a comparison of the stellar parameters within $R_{e}$/8 with these
within $R_{e}$/2 of some early-type galaxies sample,
\citet{davies,trager00,denicolo}
found that the elliptical galaxies present slightly 
negitive metallicity gradients from the centers to the outer regions
and the ages are likely to increase slightly outwards.  The same trends 
were detected by \citet{fisher}. But still only 1/3 of the star 
mass is contained within $R_e$/2. In this work, we obtained deep long slit 
spectra of the local galaxy \object{NGC4636} outer to 1.5 $R_{e}$,
aiming to study the stellar populations and kinematics out to larger redius 
and  give crucial insight into the formation paths of galaxy. 
  
The paper is organized as follows. In Sect. \ref{sec_obsdat} we
describe the observations (Sect.  \ref{sec_obs}) and the data
reduction (Sect. \ref{sec_datred}). In Sect.  \ref{sec_kinlin} we present
the kinematics (Sect. \ref{sec_kin}) and the line strength measurements
(Sect. \ref{sec_lin}). We analyze the Lick indices and derive ages,
metallicities, and $\alpha/Fe$ ratios, present the colors and mass-to-light
ratios and briefly describe the models and the method used in Sect. 
\ref{sec_stelpop}. A summary of this work is presented in Sect.
\ref{sec_summary}. The full data table of kinematics and Lick/IDS indices are 
shown in Sect.\ref{sec_app}.
 
\section{Observation and Data Reduction} 
\label{sec_obsdat} 
\subsection{Observation}
\label{sec_obs}

Long-slit spectra along major and minor axis of \object{NGC4636} were 
collected during the period of April to May in 2008 using the HET in
service mode and the Low-Resolution Spectrograph (LRS) with the E2
grism \citep{hill}. In order to detect the galaxy outer 
regions, the center of the galaxy was moved towards one end of the slit. 
The slit width was 3~arcsec, giving an instrumental
broadening of $\sigma_{inst}$ = 120 $\mathrm{km/s}$ and covering the wavelength 
range from 4790 to 5850 (\AA). The exposure time of each slit was 900 sec. 
Moreover, 900~sec exposures of blank sky regions were taken at regular 
intervals. The seeing ranged from 1.49 to 2.57 arcsec.  The resulting summed 
spectra probe regions out to nearly 1.5 $R_e$ of \object{NGC4636}.
In addition, calibration frames (biases, dome flats and the Ne and Cd
lamps) were taken. Table \ref{tab_log} shows the logs of the 
spectroscopic observations. 

 \begin{table}
 \caption{Log of spectroscopic observation, MJ = major axis; 
 MN = minor axis; SKY= sky spectrum.\label{tab_log}}
 \begin{center}
 \begin{tabular}{lllc}
 \hline
 Date       & Objects   & Position  &Seeing(FWHM)\\
 \hline

 2008 Apr 04 & \object{NGC4636} &MJ1,   SKY1    &2.57\\
 2008 Apr 06 &            &MJ2,3, SKY2    &1.49\\
 2008 Apr 09 &            &MJ4,5, SKY3     &1.64 \\
 2008 May 01 &            &MN1,2, SKY4     &1.92 \\
\hline
\end{tabular}
\begin{list}{}{}
\item[$^{\mathrm{a}}$] The exposure time for each slit is 900
seconds.
 \item[$^{\mathrm{b}}$] Position angles of major and minor axes of
 \object{NGC4636} are $150^{\circ}$ and $60^{\circ}$ respectively. 
\end{list}
\end{center}
\end{table}

\subsection{Data Reduction} 
\label{sec_datred} 

\begin{figure}
  \begin{center}
    \includegraphics[width=0.44\textwidth]{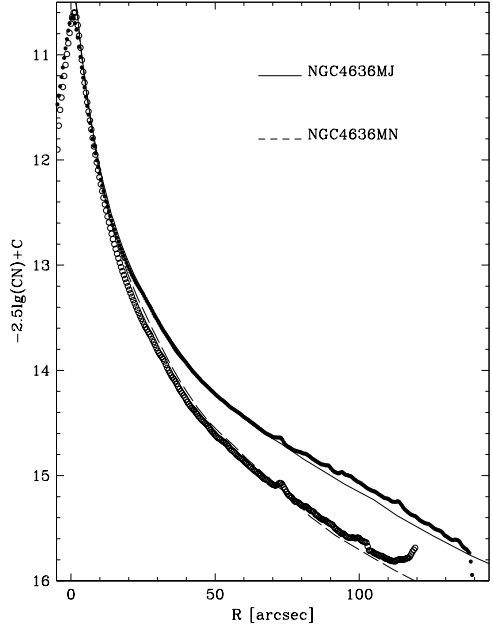}\\
  \end{center}
  \caption{The comparison between the broadband surface brightness
    profiles (lines) and the ones derived from our summed spectra (dots) 
    for \object{NGC4636}. The solid and open dots indicate the 
profiles measured from the summed spectrum along the major and minor 
axes respectively.
The solid and dashed lines present the V band photometry data taken from 
\citet{kormendy}.
\label{fig_skysubproof}}
\end{figure}
The data reduction used the MIDAS package provided by ESO.  The
pre-process of data reduction was done following
\citet{bender94}.  The raw spectra were bias subtracted,
and divided by the flatfields. The cosmic rays were removed with a
$\kappa-\sigma$ clipping procedure. The wavelength calibration was
performed using 9 to 11 strong Ne and Cd emission lines and a third
order polynomial. The achieved accuracy of the wavelength calibration
is always better than 0.6 {\AA} (rms).  The science spectra were
rebinned to a logarithmic wavelength scale. 

The step of sky subtraction required particular care to minimize
systematic effects on the measured kinematics and line strengths in
the outer regions of our galaxies. More detials can be seen in 
\citet[ Fig. 2]{saglia10} and \citet[Fig.1;Fig2]{pu2010}. 
Here, We briefly describe the calibration of the atmospheric sky level 
procedure.  At the begining, we selected the spectra of galaxy
where a sky spectrum with a uniform slit illumination was available, and
almost photometric conditions were achieved, yielding the largest galaxy
counts per pixel. To correct for the inhomogeneous slit illumination, 
we produced a 4th to 6th order polynomial model of the sky spectra for 
each column in spatial direction and subtracted it from the selected galaxy frames,
obtaining a reference science frames $G_r$. We computed the fractional residuals 
between the scaled and the reference slit profiles
\begin{equation}
\label{eq_sky}
R(r)=|1-\frac{f_i^G \ast \langle G_r\rangle}{\langle (G_i -f_i^S\ast SKY_i)\rangle}|.
\end{equation}
and minimized it (see below).
Here $f_i^S$ is the scaling factor of the noise-free (i.e. the
polynomial model) sky frame taken after the galaxy frame $G_i$, when
available, or the average of the most uniform sky frames when not.
The symbol $\langle\rangle$ indicates the average in the wavelength
direction and $R(r)$ is a function of the position $r$ along the slit.
Moreover, $f_r^G$ is a scaling factor that takes into account the
different atmospheric transmissions.  We determined $f_i^G$ and $f_i^S$
iteratively such to minimize R, which in an ideal situation should be
zero at every radii. Finally, we computed the resulting total galaxy
frame $G_{tot}$ as:
\begin{equation}
\label{eq_tot}
G_{tot}=G_r+\Sigma_i (G_i-f_i^S \ast SKY_i)/f_i^G.
\end{equation}
In practice, due to the non-uniformity of the slit illumination
function the function $R(r)$ is not always zero, but through the
summing in Eq. \ref{eq_tot} the differences should average out. We can
test the quality of the calibration by comparing the profile $\langle
G_{tot}\rangle$ with available broadband photometry.
 Figure \ref{fig_skysubproof} shows the comparison between the
summed shifted slit profiles and the broadband photometry of the 
galaxy. The solid and open dots indicate the counts number 
profiles measured from the summed spectrum of major and minor axes respectively. 
The solid and dashed lines present the V band photometry data taken from 
\citet{kormendy}. The plot confirm that the summed spectrum agrees well 
with the broadband photometry out to large radius. 

In addition, we also need to correct the anamorphic distortion of the LRS 
, remove sky emission line spectrum and remove the continummu spectra. 
The procedures are described in details in \citet{pu2010}.
%**********************************************
\section{Kinematic and Lick/IDS indices Profiles} 
\label{sec_kinlin} 
\subsection{Kinematics Profiles}
\label{sec_kin}
\begin{figure}
 \begin{center}
  \begin{tabular}{c}
 \includegraphics[width=0.42\textwidth]{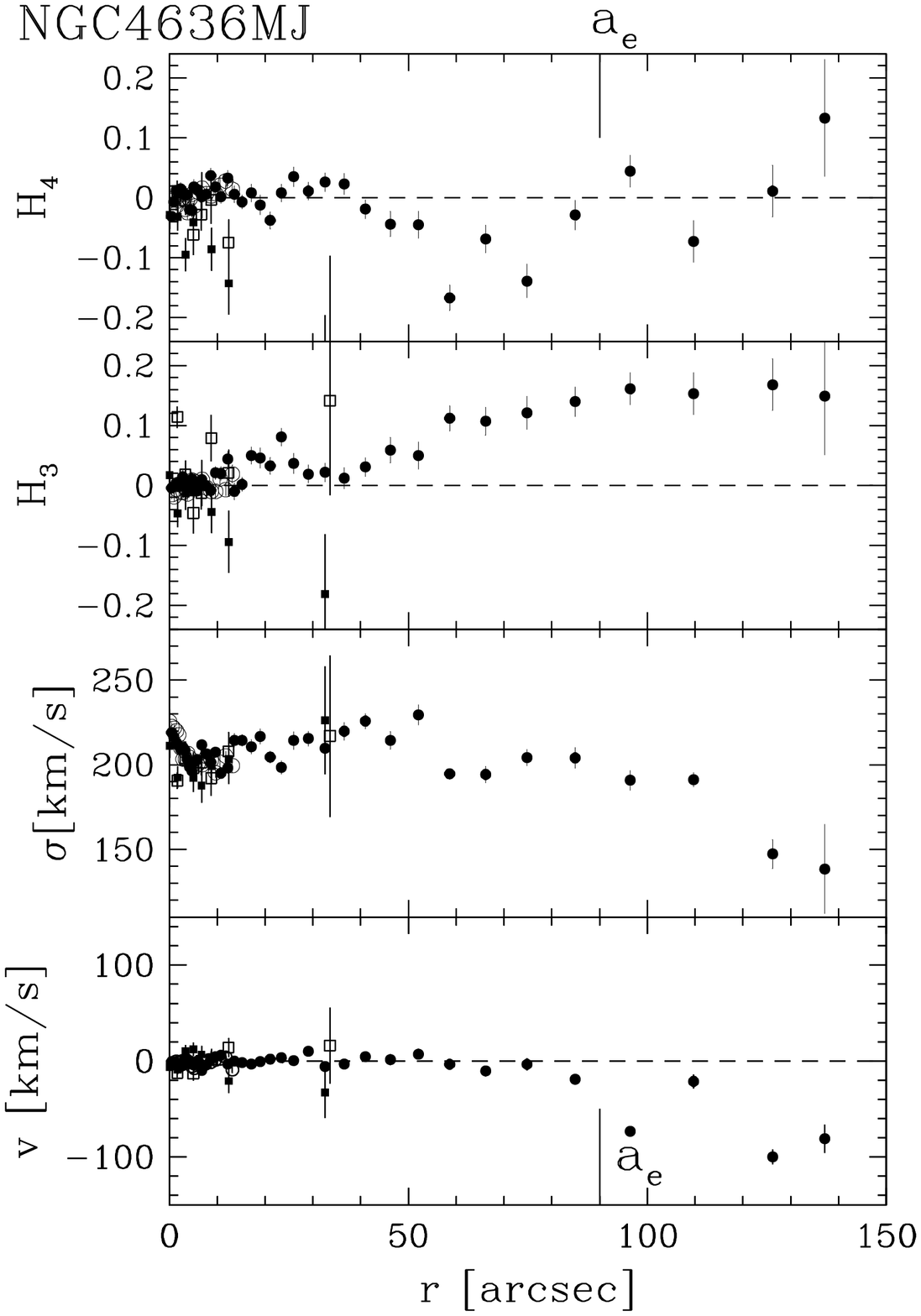}
 \includegraphics[width=0.4\textwidth]{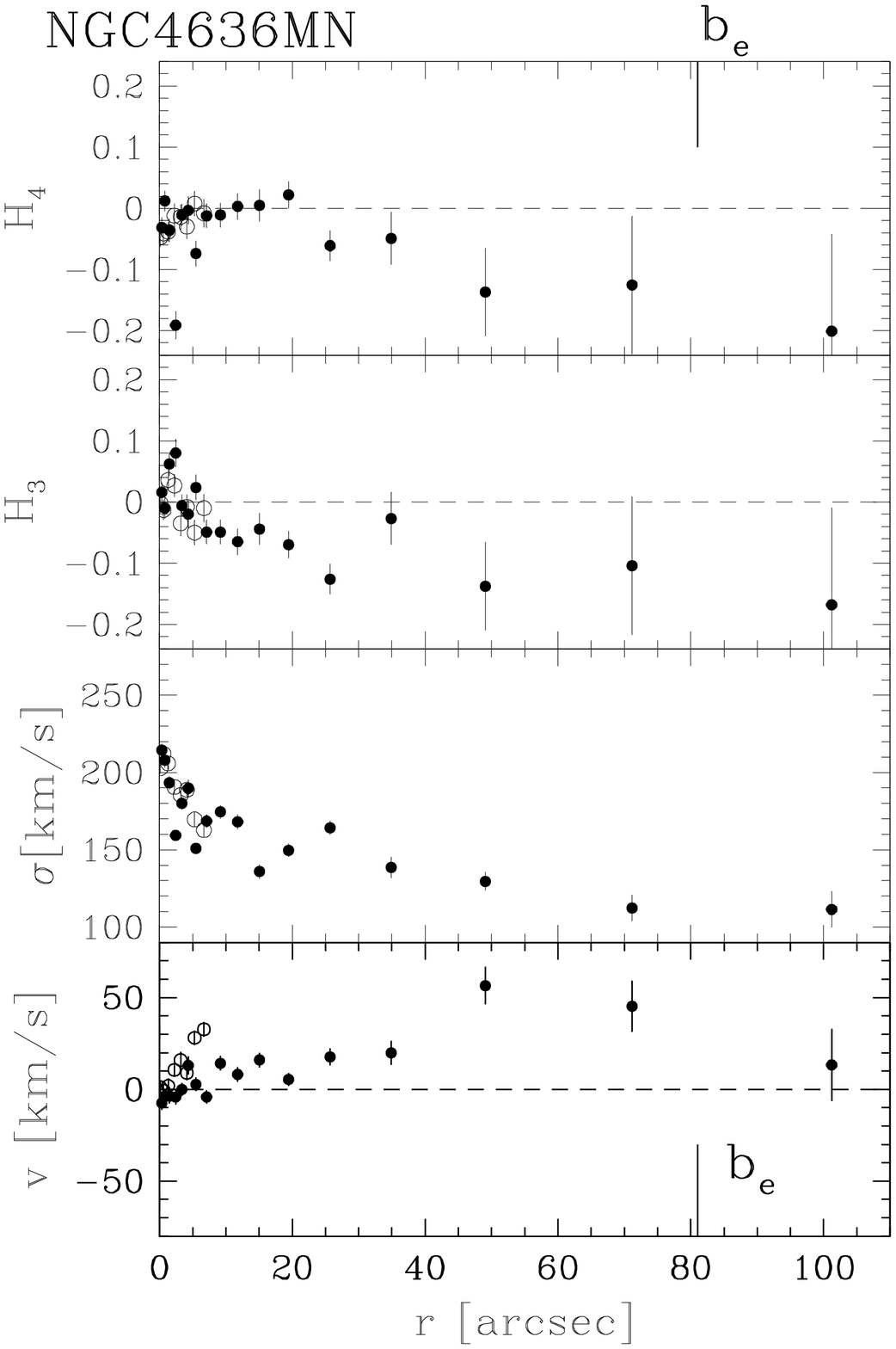} 
  \end{tabular}
 \end{center}
 \caption{The stellar kinematics profiles along the major and 
minor axes of \object{NGC4636}. From bottom to top in each panel 
we show: (1) Rotation velocity, (2) Velocity dispersion, (3) and (4)
Gauss-Hermite parameters $H_3$ and $H_4$. The profiles are folded with respect 
to the nucleus of the galaxis, filled and open symbols stand for different sides 
of the galaxy. The squares show the data published in \citet{bender94}.
\label{fig_kin}}
 \end{figure}
We extracted the line-of-sight velocity distributions (LOSVDs) and kinematic
parameters from the continuum-removed spectra that rebinned radially to obtain 
almost constant signal-to-noise ratio,
using the Fourier Correlation Quotient (FCQ) method
\citep{bender90} with the implementation described in
\citet{saglia10} that allows for the presence of emission
lines. The stellar spectra library of \citet{vazdekis} is used 
as the templates aimming to minimize the mismatching. This library contains 
about a thousand synthetic single-stellar-population spectra covering the 
wavelength range from 4800 to 5470 {\AA} with a resolution of 1.8 {\AA}. 
We used the library with ages of 1.00 to 17.78 Gyr and 
metallicities from -1.68 to 0.2. We first set all of the library
spectra to the resolution of our galaxy spectra and find the best
fitting template for each radial bin according to the lowest RMS value
of the residual (reaching typically  1\% of the initial flux). 
If emission lines are detected, Gaussians are fitted
to the residuals above the best-fit template and subtracted from the
galaxy spectrum to derive cleaned spectra. The kinematic fit is then
redone using these cleaned spectra.  We do not detect emission in the
spectra of \object{NGC4636}.

The Fig. \ref{fig_kin} presents the kinematics along the major  
and minor axes in \object{NGC4636}. In this figure we
show the rotational velocity, the velocity dispersion and the
Gauss-Hermite parameters $H_3$ and $H_4$. The filled symbols
show the kinematic profiles on south-east (SE) side along 
the major axis and north-east (NE) side along the 
minor axis. The reference data taken from \citet{bender94} 
are also doted in squares in the plot. As it can be seen from the
figures, agreement is generally good. The $a_e = R_e \cdot \epsilon^{-1/2}$ and 
$b_e = R_e \cdot \epsilon^{1/2}$ are labeled, where the R$_e$ is the effective 
radius and the $\epsilon$ is the apparent axial ratio. The kinematics data 
extend to 140 arcsec along major axis and 105 along minor axis from the galaxy 
center. Further extend out than the previous work of \citet{bender94,
davies,rampazzo}. The kinematic profiles present no 
rotation both along major and minor axes. The velocity dispersion show flat 
gradients inside 100 arcsec and become steep along major axis. The measured 
stellar kinematics with errors table is presented in 
Table.\ref{tab_kine} in Appendix.

 %********************************************************************
\subsection{Lick/IDS indices Profiles} 
\label{sec_lin}
\begin{figure*}
 \begin{center}
 \begin{tabular}{c}
 \includegraphics[scale=0.42]{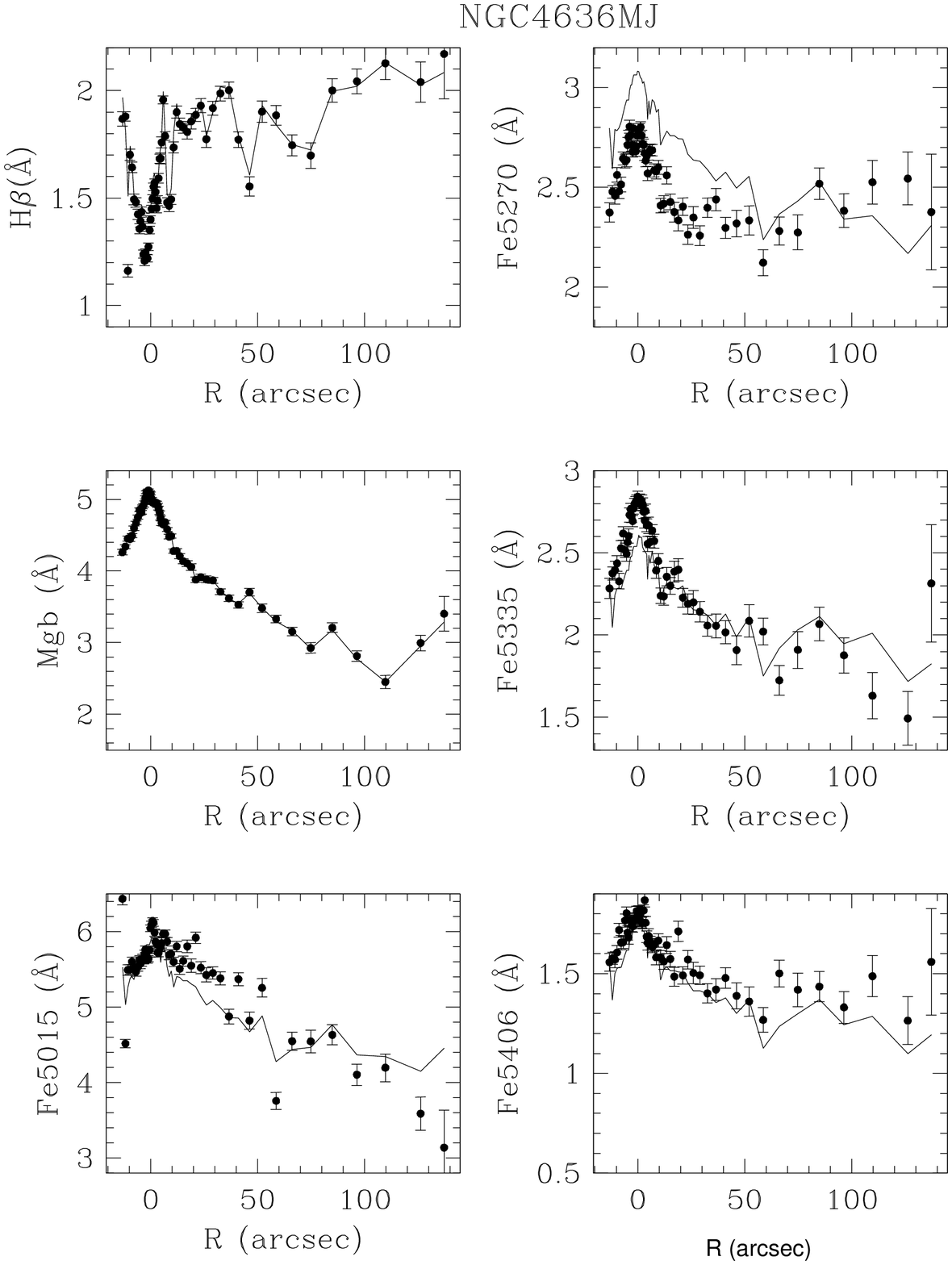}
 \includegraphics[scale=0.42]{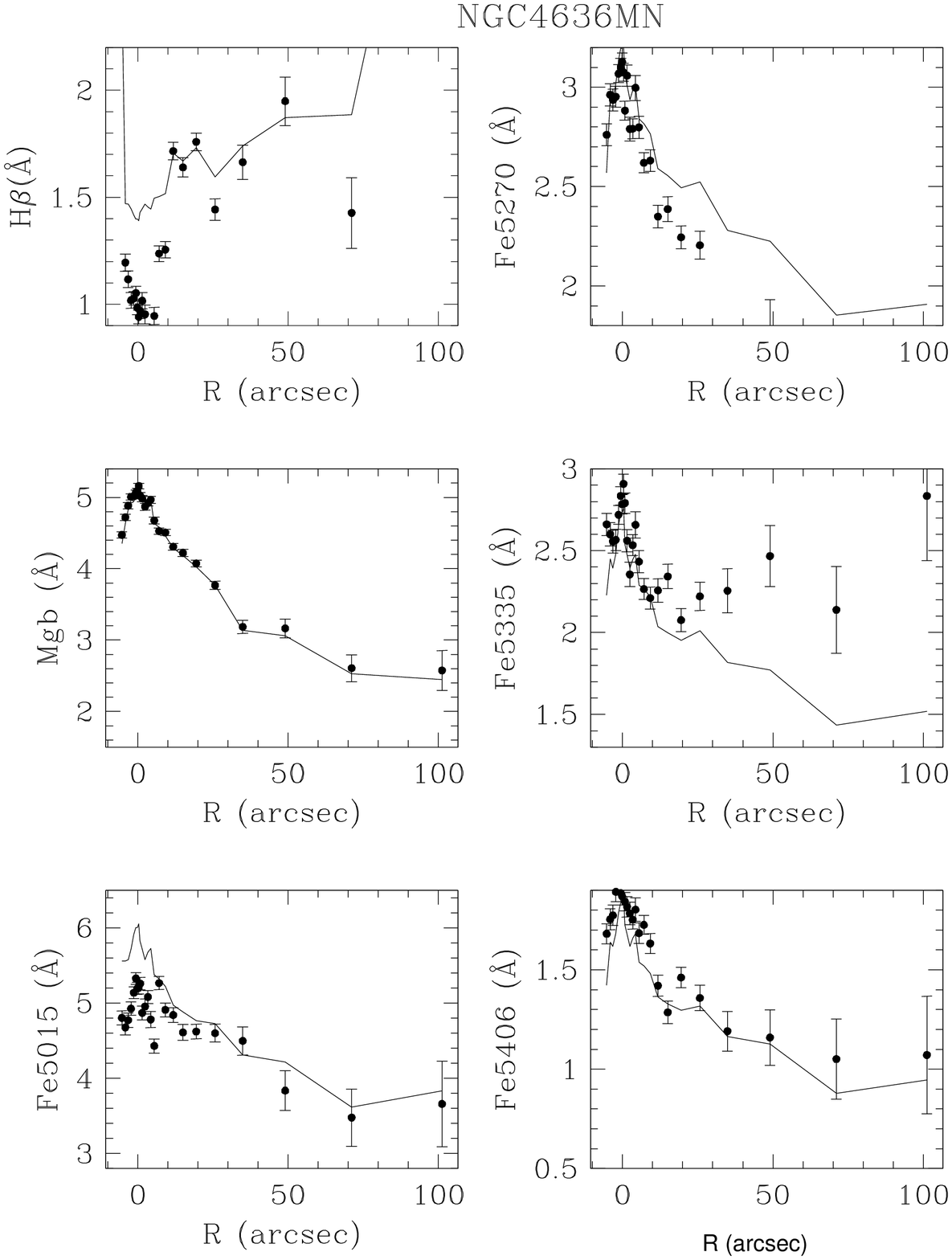}
 \end{tabular}
 \end{center}
 \caption{The line-strength indices along the major and minor
   axis. The names of the indices are labeled
   and the galaxies' names are also noted. The solid lines show the
   model (TMB03) predicted line strength profiles along the axes.
   \label{fig_lin}}
 \end{figure*}
 
 In this section, we describe the measurement of six Lick/IDS indices 
($H\beta,Mgb,Fe_{5015},Fe_{5270},Fe_{5335},Fe_{5406}$), which defined by 
\citet{trager98}. The line strength indices have been measured 
from the cleaned spectra along the major and minor axis. Before measuring the
indices, our spectra were degraded to the resolution of Lick/IDS
systems. We then corrected the indices for the velocity dispersion
using template stars and the value for $\sigma$ derived in the
previous section. Finally, the observational data need to be corrected
to the Lick/IDS system.  To do this we observed 5 stars from the
Lick/IDS library using the same instrumental configuration used for
the science objects and derived the offsets between our data and the
Lick/IDS system.  The comparison of our data with the Lick standard
systems can be found in \citet[Fig. 4]{pu2010}. The data are in good agreement 
with the Lick/IDS systems, as found in \citet{saglia10} using a 
larger set of Lick standards observed with LRS and HET at a better resolution. 
So far, the deviation between our measurements and the Lick system can 
be ignored, but we take into account the RMS of the calibration lines into the 
final error budget, by adding it in quadrature to the statistical error of each index.
Fig.\ref{fig_lin} shows the six line strength indices profiles along the major 
and minor axies of the galaxy. The names and positions are labeled in 
the figure. The dots present the measured lick/IDS and the solid lines show the SSP 
models predictions, they will be discussed in the following sections. 
The full table of six Lick/IDS indices is presented in Table. 
\ref{tab_lick} in the Appendix.
The galaxy presents positive gradients of H$\beta$ index both along major axis and 
minor axis, very similar to profiles in another galaxies discovered in 
previous work \citep{davies,sanchez07,pu2010}. We also measured 
the indices $Mg_1$ and $Mg_2$, but we do not use them in this work since these two 
indices are very sensitive to the anamorphic distortion.

%************************************************************************
\section{Stellar Populations}
\label{sec_stelpop}
In this section, we derive the age, total metallicity and element
abundance gradients along the major and minor axes by comparing the 
measured line indices with simple stellar population models
(TMB03). The details of this model can be seen from 
\citet{thomas03,thomas05}. Here, we give a 
brief description of this model: The TMB03 models cover ages
between 1 and 15 Gyr, metallicities between 1/200 and 3.5 solar.
Furthermore, the models take into account the effects on the Lick
indices by the variation of $\alpha$ abundance, hence, give Lick
indices of simple stellar populations not only as the function of age
and metallicity, but also as the function of the $\alpha/Fe$ ratio. 

In this work, we do not adopt the traditional and effective method of 
studying stellar population properties which uses diagrams of
different pairs of Lick indices \citep{thomas05}.  
The method selects the $H\beta$ versus $[MgFe]^{\prime}$ pair
diagram to break the age-metallicity degeneracy, where, $[MgFe]^{\prime}$
=$\sqrt{Mgb[0.72 Fe5270+0.28 Fe5335]}$, because $H\beta$ is
sensitive to warm turnoff stars and $[MgFe]^{\prime}$ index is considered as
the best detector of metallicity since it does not depend on abundance
ratio variations. Following
\citet{saglia10} and \citet{pu2010}, we use the simple $\chi^{2}$
minimization method. $\chi^{2}$ method fits all of  $H\beta$, Mgb, 
Fe5270, Fe5335 and other available indices at the same time, 
the best resolution is bound to 
break the age-metallicity degeneracy. The tests for Coma galaxies
 done by \citet{thomas11} show that the results which derived 
using new method were better than the results obtained using the traditional method.
 The $\chi^{2}$ is give by:   
\begin{equation}
\label{eq_chi}
\chi^{2}=\sum_{index[i]}\frac{(index_{\rm i[ob]}-index_{\rm i[mod]})^{2}}
{(\sigma_{\rm i[ob]})^{2}}
\end{equation}
where $index_{\rm i[ob]}$ and $index_{\rm i[mod]}$ represent the
$i^{th}$ observational indices and model indices respectively,
$\sigma_{\rm i[ob]}$ is the observational uncertainty of $i^{th}$
indices. We can derive the best fitting age, metallicity and 
$\alpha/Fe$ by finding the minimum $\chi^{2}$ of all selected lines
indices to the SSP models. The H$\beta$, Mg\emph{b},
$Fe_{5015}$, $Fe_{5270}$, $Fe_{5335}$ and $Fe_{5406}$ are used as the
indicators. Moreover, we interpolated the tabulated
indices of TMB03 on steps of 0.1 Gyr in age, 0.02 in
metallicity and 0.05 in $\alpha/Fe$ aimming to improve the precision 
of the stellar properties using the $\chi^{2}$ minimization method.

\begin{figure}
\begin{center}
%\begin{tabular}{c}
\includegraphics[width=0.44\textwidth]{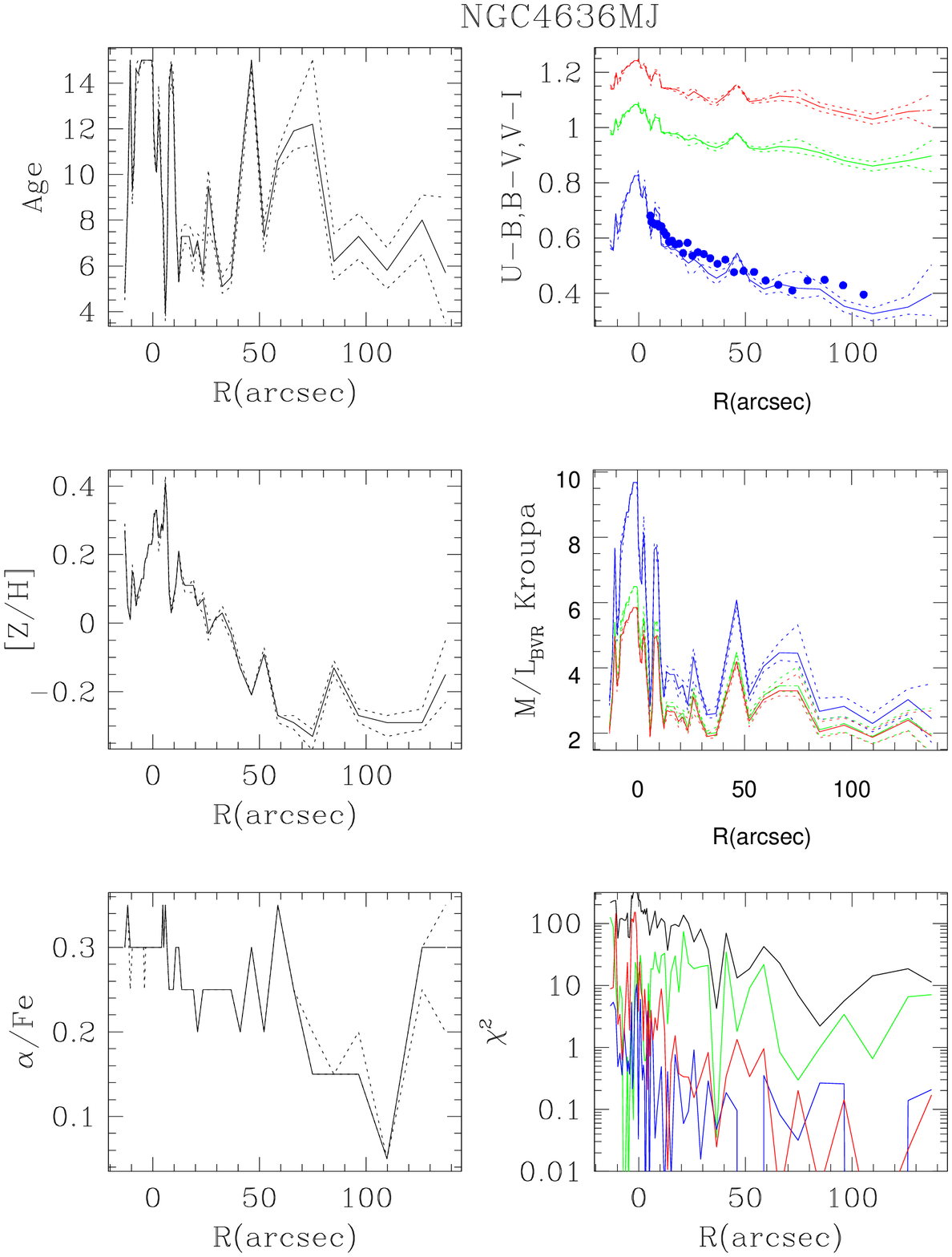}
\includegraphics[width=0.44\textwidth]{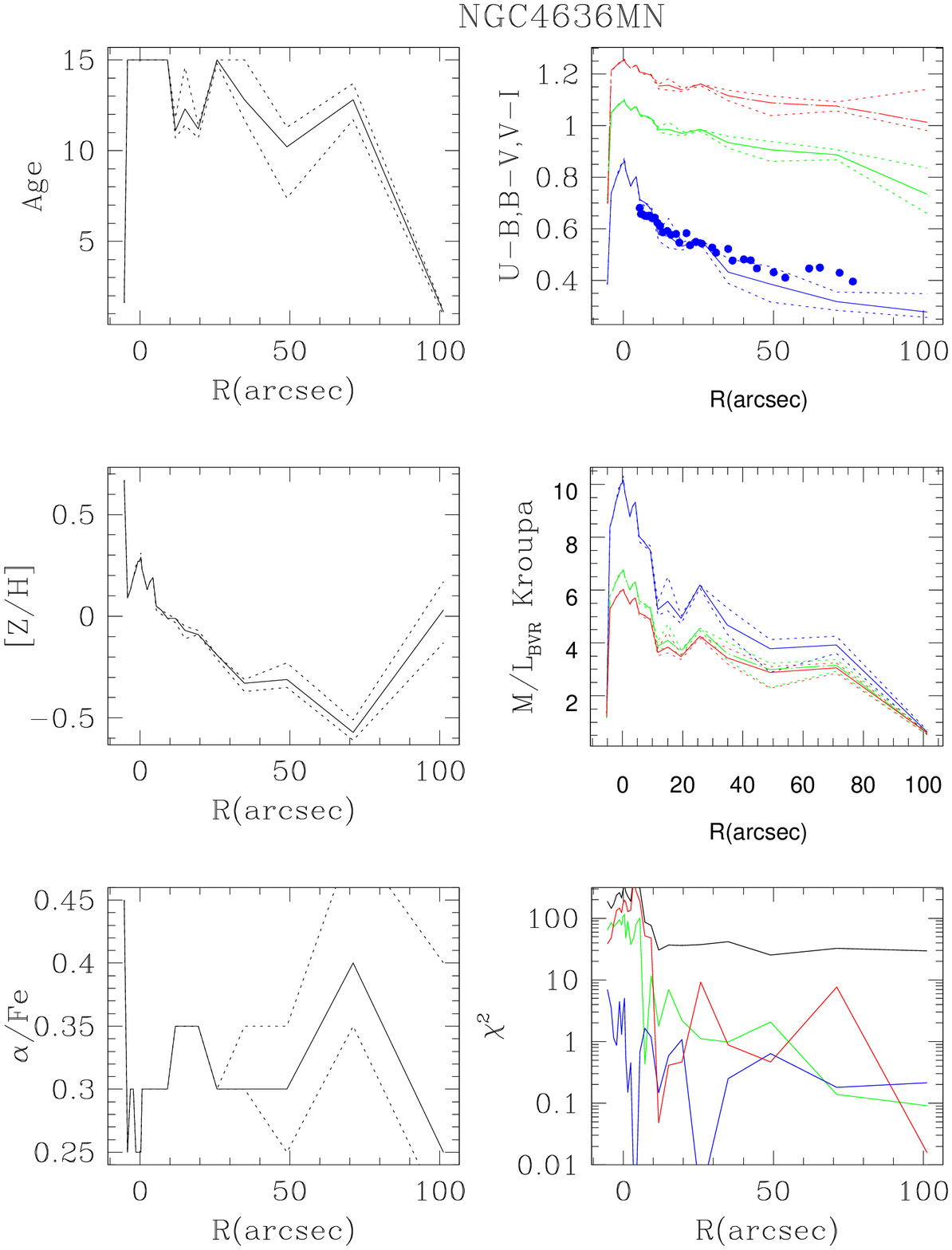}
%\end{tabular}
 \end{center}
 \caption{The best fitting SSP equivalent age, metallicity and
   element abundance ratio are shown in the left in each
   plot. The right top of each plot shows the Johnson
   broadband U-B, B-V, V-I colors profiles, the blue and green
   solid lines stand for U-B and B-V color respectively;
   the red solid lines indicate the V-I color. The measured U-B color 
   taken from \citet{peletier} are doted with blue solid
   symbols. The M/L$_{BVI}$ are shown in the mid of the right columns, 
   the blue, green and red solid lines indicate the M/L in the B, V
   and I band respectively; the minimized $\chi^{2}$ of selected line 
   strengths are presented in the bottom panel on the right, red, 
   blue and green lines 
   present the minimized $\chi^{2}$ of H$\beta$, Mg\emph{b} and Fe$_{5270}$ 
   respectively; while the black lines show the total minimized $\chi^{2}$. 
   \label{fig_SSP}}
 \end{figure}

Fig. \ref{fig_SSP} presents the model predicted age, metallicity, 
element abundance, colors, mass-to-light ratios, and resulting
$\chi^2$. From the top left panel to the bottom left panel, the age, 
total metallicity [Z/H] and $\alpha/Fe$ ratio are shown. The galaxy presents 
$\alpha$ overabundant from the galaxy center to the outer parts. The total 
metallicity profile shows a steep negative gradients with $\triangle [Z/H]/\triangle 
\log(r)$ = -0.333 $\pm{0.022}$ inner $R_e$ and become flat outwards.  
The average age, metallicity and $\alpha/Fe$ inside 1/16
$R_{e}$ are 9.96 $\pm{1.83}$ Gyr, 0.302 $\pm{0.08}$ and 0.325 $\pm{0.025}$. 
They are comparable to the results of age = 9.918 $\pm{0.458}$ derived 
by \citet{sanchez06} using the $H\beta$ versus [Mgb] pair
diagram and the age = 8.18 $\pm{0.063}$ in \citet{proctor}.  
The model predicted lines strength profiles are shown in Fig.
\ref{fig_lin} with solid lines. As it can be seen from
Fig. \ref{fig_lin}, in generally, theoretical line strength indices 
match well with the measured parameters in the inner regions of the
galaxies except Fe5270 due to this indies be contaminated by some unkonwn
sky lines and we set its weight to zero. In addition, we also calculate the 
Johnson broadband U-B, U-V, B-V, V-R, V-I, V-K, J-K, J-H, H-K color and M/L
ratio in B, V, R, I, J, H, and K bands profiles using the Kroupa initial
mass function \citep{kroupa} with the help of the SSP models
\citep{maraston}. For a clear presentation in the figures,
we only show the U-B, B-V and V-I color and the mass to light ratios
in B, V, I band in this papers. On the top right panels in Fig.
\ref{fig_SSP}, the blue and green solid lines stand for U-B and B-V
color respectively; the red lines indicate the V-I color. The measured
U-B color [taken from \citep{peletier}] are also over 
plotted with solid blue dots. As it can be seen from the plot, the model
predicted color profiles agree well with the measured colors. Indeed, the 
colors present steep negative gradients and sharp peaks, this feature mainly 
shaped by the metallicity profile. The middle right panel of Fig.\ref{fig_SSP} shows 
the theoretical M/L ratio in B, V, I bands; the blue, green and red lines 
display the M/L ratios in B, V and I colors respectively; The minimized 
$\chi^{2}$ of selected line strength are presented in the bottom panel on 
right, red, blue and green lines present the minimize $\chi^{2}$ of H$\beta$, 
Mg\emph{b} and Fe$_{5270}$ respectively; while the black lines show the total
minimized $\chi^{2}$. The large values of $\chi^{2}$ along major axis 
are mainly driven by the Fe5270.

%***************************************************************************
\section{Sumarry and Discussion}
\label{sec_summary}
In this work, we measured the accurate kinematic profiles extending
out to 1.5 $R_e$ along the major and minor axis for the giant
elliptical galaxy \object{NGC4636}. Indeed, six Lick line indices 
($H\beta,Mgb,Fe_{5015},Fe_{5270},Fe_{5335},
   Fe_{5406}$) defined by \citet{burstein84,
worthey,trager98} of \object{NGC4636} 
are also derived. By comparing the measured Lick/IDS with the SSP
model predictions, we derived the stellar population
parameters, the M/L ratios and the broadband colors of our
galaxies. We found the galaxy \object{NGC4636} has high
metallicities in the galaxy center and presents steep negative metallicity
gradients. The galaxy is $\alpha$ overabundant and do not present significant 
gradients of $\alpha$ abundances along the axes.  The galaxy have sharp 
red peaks at the center which be mainly shaped by the metallicity 
\citep{maraston}. The model predicted colors agree well with
the measured color profiles.

According to the simple element enrichment scenario, the $\alpha$
elements are mainly delivered by Type II supernovae explosions of
massive progenitor stars, and a substantial fraction of Fe peak
elements come from the delayed exploding Type Ia supernovae
\citep{nomoto,thielemann}. Thus the $\alpha/Fe$
can be used as an indicator to constrain the formation timescale of
stars. Hence the flat profile of $\alpha$/Fe ratio in
our galaxy likely suggests that there is no radial variation of star
formation time scales.  The radial metallicity and lines strength
gradients give one of the most stringent constraints on the galaxy
formation \citep{sanchez06,tortora}. 
The galaxies that form monolithically have steeper
gradients and the galaxies that undergo major mergers have shallower
gradients.  The mean metallicity gradients for non-merger and merger
galaxies derived by theoretical simulation in
\citet{kobayashi} are $\triangle$[Z/H]/$\triangle$ log(r)
$\sim -0.30 \pm 0.2$ and $-0.22 \pm 0.2$, respectively. The author
found that the galaxies with gradients steeper than -0.35 are all
non-major merger galaxies.  The gradients inside $R_{e}$ of \object{NGC4636} is
$\triangle [Z/H]/\triangle \log(r)$ = -0.333
$\pm{0.022}$; while the gradients of \object{NGC4636}
becomes flat outer $R_e$. Accordingly, this is a weak indication that
the center and the out regions of \object{NGC4636} are formed through
different formation process. Indeed, it is worthwhile to observe the 
Lick/IDS in further out regions with newly developed observing technique 
as done by \citet{weijmans}. And it will give us a 
new insight on the formation of the elliptical galaxies. 

In the forthcoming works, we plan to further investigate the
dynamical structure and orbit distribution of \object{NGC4636}, 
to investigate further constraints on the formation process of the galaxy.  

\begin{acknowledgements}
We specially thank the McDonald Observatory for performing the
observations with Hobby-Eberly Telescope (HET) in service mode.  The
HET is a joint project of the University of Texas at Austin, the
Pennsylvania State University, Stanford University,
Ludwig-Maximilians-Universit\"{a}t M\"{u}nchen, and
Georg-August-Universit\"{a}t G\"{o}ttingen. The HET is named in
honor of its principal benefactors, William P. Hobby and Robert E.
Eberly." The Marcario Low Resolution Spectrograph is named for Mike
Marcario of High Lonesome Optics who fabricated several optics for
the instrument but died before its completion. The LRS is a joint
project of the Hobby-Eberly Telescope partnership and the Instituto
de Astronom\'{\i}a de la Universidad Nacional Aut\'{o}noma de
M\'{e}xico. This work was in part supported by the Chinese National
Science Foundation (Grant No. 10821061) and the National Basic
Research Program of China (Grant No. 11033008, 2007CB815406). We also
gratefully acknowledge the Chinese Academy of Sciences and
Max-Planck-Institut f\"{u}r extraterrestrische Physik that
partially supported this work. Z.Han thanks the support of the
Chinese Academy of Sciences (Grant No. KJCX2-YW-T24). We thank
 Prof. Bender gave us the HET time to perform the obserations. 
We are grateful to Dr. R.P.Saglia and J.Thomas for reading the manuscript 
and useful suggestion. 
We really appreciate the referee for useful comments that help us to 
improve the presentation of the results.
\end{acknowledgements}

%\appendix
\section{Appendix}
\label{sec_app}
\begin{table*}[h]
\caption{The full table of measured stellar kinematics as a function of distance from the center 
(positive: Southeast, negative: Northwest for the position angle of $150^{\circ}$;
 and positive: Notheast, negatice: Southwest for the position angle of $60^{\circ}$)}.
\label{tab_kine}
\begin{tabular}{lcrccccccccc}
\hline
Galaxy & PA& R &V&$\pm$dV&$\sigma$&$\pm$d$\sigma$&$H_3$&$\pm$d$H_3$&$H_4$&$\pm$d$H_4$ &S/N  \\ 
Name   & (deg) & (\arcsec) & (km/s) & (km/s)& (km/s) & (km/s) & & & & &\\
\hline
 \object{NGC4636}&150&-13.17&-9.18&$\pm$3.36&199.4&$\pm$4.11& 0.019&$\pm$0.015& 0.014&$\pm$0.015 &121.4\\ 
 \object{NGC4636}&150&-11.77& 3.97&$\pm$2.27&205.2&$\pm$2.87&-0.007&$\pm$0.010& 0.028&$\pm$0.010 &185.3\\
 \object{NGC4636}&150&-10.60& 2.26&$\pm$2.92&201.6&$\pm$3.64& 0.011&$\pm$0.013& 0.021&$\pm$0.013 &141.2\\
 \object{NGC4636}&150&-9.66 & 1.18&$\pm$2.55&195.3&$\pm$3.07&-0.010&$\pm$0.012& 0.007&$\pm$0.012 &156.8\\
 \object{NGC4636}&150&-8.72 &-1.21&$\pm$2.64&202.8&$\pm$3.13&-0.010&$\pm$0.012& 0.003&$\pm$0.012 &157.7\\
 \object{NGC4636}&150&-7.78 &-0.25&$\pm$2.66&205.1&$\pm$3.20&-0.004&$\pm$0.012& 0.007&$\pm$0.012 &157.9\\
 \object{NGC4636}&150&-6.84 &-0.67&$\pm$2.27&200.0&$\pm$2.79& 0.010&$\pm$0.010& 0.016&$\pm$0.010 &180.2\\
 \object{NGC4636}&150&-5.90 &-2.00&$\pm$2.40&200.2&$\pm$2.91& 0.000&$\pm$0.011& 0.010&$\pm$0.011 &170.7\\
 \object{NGC4636}&150&-5.21 &-7.68&$\pm$2.39&200.9&$\pm$2.72& 0.007&$\pm$0.011&-0.014&$\pm$0.011 &172.5\\
 \object{NGC4636}&150&-4.74 &-5.64&$\pm$2.75&201.4&$\pm$3.29& 0.009&$\pm$0.012& 0.006&$\pm$0.012 &150.1\\
 \object{NGC4636}&150&-4.27 &-1.67&$\pm$2.48&204.2&$\pm$2.91& 0.000&$\pm$0.011&-0.003&$\pm$0.011 &168.5\\
 \object{NGC4636}&150&-3.80 &-0.92&$\pm$2.41&206.6&$\pm$2.67&-0.014&$\pm$0.011&-0.025&$\pm$0.011 &175.5\\
 \object{NGC4636}&150&-3.33 &-0.78&$\pm$1.99&203.4&$\pm$2.25&-0.003&$\pm$0.009&-0.019&$\pm$0.009 &208.9\\
 \object{NGC4636}&150&-2.86 & 0.47&$\pm$2.28&203.3&$\pm$2.70&-0.011&$\pm$0.010& 0.000&$\pm$0.010 &182.7\\
 \object{NGC4636}&150&-2.39 &-1.07&$\pm$2.03&209.9&$\pm$2.43&-0.012&$\pm$0.009& 0.006&$\pm$0.009 &212.0\\
 \object{NGC4636}&150&-1.92 &-1.35&$\pm$1.93&217.7&$\pm$2.25& 0.002&$\pm$0.008&-0.005&$\pm$0.008 &231.5\\
 \object{NGC4636}&150&-1.45 &-0.03&$\pm$1.91&220.0&$\pm$2.17&-0.006&$\pm$0.008&-0.014&$\pm$0.008 &236.2\\
 \object{NGC4636}&150&-0.98 &-2.39&$\pm$1.96&221.5&$\pm$2.31&-0.019&$\pm$0.008&-0.001&$\pm$0.008 &231.3\\
 \object{NGC4636}&150&-0.51 &-2.28&$\pm$2.01&222.5&$\pm$2.41&-0.015&$\pm$0.008& 0.007&$\pm$0.008 &227.3\\
 \object{NGC4636}&150&-0.04 &-2.14&$\pm$2.08&225.6&$\pm$2.42& 0.002&$\pm$0.008&-0.005&$\pm$0.008 &222.5\\
 \object{NGC4636}&150& 0.44 &-0.68&$\pm$1.90&219.1&$\pm$2.08&-0.004&$\pm$0.008&-0.031&$\pm$0.008 &235.6\\
 \object{NGC4636}&150& 0.91 &-3.11&$\pm$1.88&216.5&$\pm$2.18&-0.002&$\pm$0.008&-0.007&$\pm$0.008 &235.7\\
 \object{NGC4636}&150& 1.38 &-2.54&$\pm$1.98&213.3&$\pm$2.41& 0.004&$\pm$0.008& 0.012&$\pm$0.008 &220.5\\
 \object{NGC4636}&150& 1.85 &-2.33&$\pm$1.99&211.4&$\pm$2.40&-0.002&$\pm$0.009& 0.008&$\pm$0.009 &217.4\\
 \object{NGC4636}&150& 2.32 &-5.90&$\pm$2.28&209.1&$\pm$2.79& 0.007&$\pm$0.010& 0.015&$\pm$0.010 &188.0\\
 \object{NGC4636}&150& 2.79 &-4.76&$\pm$2.23&210.2&$\pm$2.70& 0.015&$\pm$0.010& 0.010&$\pm$0.010 &192.9\\
 \object{NGC4636}&150& 3.26 & 1.75&$\pm$2.59&208.6&$\pm$3.09&-0.003&$\pm$0.011& 0.004&$\pm$0.011 &164.9\\
 \object{NGC4636}&150& 3.73 & 1.93&$\pm$2.24&203.2&$\pm$2.68&-0.005&$\pm$0.010& 0.005&$\pm$0.010 &185.7\\
  \hline
\end{tabular}
\end{table*}

\addtocounter{table}{-1}
\begin{table*}
\caption{Continued} 
\begin{tabular}{lcrccccccccc}
\hline
Galaxy & PA& R &  V &$\pm$ dV  &  $\sigma$& $\pm$ d$\sigma$  &  $H_3$  &$\pm$ d$H_3$  & $H_4$ &$\pm$ d$H_4$  & S/N\\ 
Name   & (deg) & (\arcsec) & (km/s) & (km/s)& (km/s) & (km/s) & & & & &\\
\hline 

 \object{NGC4636}&150& 4.20 & 0.42&$\pm$2.21&198.8&$\pm$2.49& 0.003&$\pm$0.010&-0.020&$\pm$0.010 &183.9\\
 \object{NGC4636}&150& 4.67 &-1.36&$\pm$2.54&196.4&$\pm$2.84& 0.011&$\pm$0.012&-0.021&$\pm$0.012 &158.4\\
 \object{NGC4636}&150& 5.14 &-4.15&$\pm$2.72&201.7&$\pm$3.35& 0.006&$\pm$0.012& 0.018&$\pm$0.012 &152.2\\
 \object{NGC4636}&150& 5.83 &-0.15&$\pm$2.01&203.7&$\pm$2.45&-0.009&$\pm$0.009& 0.013&$\pm$0.009 &207.3\\
 \object{NGC4636}&150& 6.77 &-9.65&$\pm$2.14&211.7&$\pm$2.54& 0.010&$\pm$0.009& 0.002&$\pm$0.009 &202.3\\
 \object{NGC4636}&150& 7.71 &-4.19&$\pm$2.37&206.3&$\pm$2.85&-0.001&$\pm$0.010& 0.006&$\pm$0.010 &178.1\\
 \object{NGC4636}&150& 8.65 & 1.46&$\pm$2.73&201.5&$\pm$3.52&-0.008&$\pm$0.012& 0.037&$\pm$0.012 &151.3\\
 \object{NGC4636}&150& 9.59 & 4.50&$\pm$2.46&207.4&$\pm$3.04& 0.021&$\pm$0.011& 0.018&$\pm$0.011 &172.5\\
 \object{NGC4636}&150&10.76 & 6.03&$\pm$2.50&195.1&$\pm$2.97& 0.020&$\pm$0.012& 0.002&$\pm$0.012 &159.6\\
 \object{NGC4636}&150&12.17 &-3.08&$\pm$2.84&198.2&$\pm$3.63& 0.044&$\pm$0.013& 0.033&$\pm$0.013 &142.9\\
 \object{NGC4636}&150&13.58 & 0.00&$\pm$3.22&214.5&$\pm$3.87&-0.010&$\pm$0.014& 0.006&$\pm$0.014 &136.3\\
 \object{NGC4636}&150&15.22 &-1.65&$\pm$2.95&214.3&$\pm$3.42& 0.002&$\pm$0.012&-0.007&$\pm$0.012 &149.0\\
 \object{NGC4636}&150&17.10 &-3.33&$\pm$3.58&210.7&$\pm$4.32& 0.050&$\pm$0.015& 0.008&$\pm$0.015 &120.6\\
 \object{NGC4636}&150&18.98 &-0.47&$\pm$3.98&216.8&$\pm$4.56& 0.046&$\pm$0.017&-0.012&$\pm$0.017 &111.6\\
 \object{NGC4636}&150&21.09 & 2.07&$\pm$3.21&204.6&$\pm$3.44& 0.033&$\pm$0.014&-0.038&$\pm$0.014 &130.4\\
 \object{NGC4636}&150&23.44 & 3.28&$\pm$3.34&198.5&$\pm$4.03& 0.081&$\pm$0.015& 0.008&$\pm$0.015 &121.7\\
 \object{NGC4636}&150&26.02 & 0.60&$\pm$4.12&214.4&$\pm$5.30& 0.037&$\pm$0.017& 0.035&$\pm$0.017 &106.5\\
 \object{NGC4636}&150&29.07 & 9.97&$\pm$3.61&215.5&$\pm$4.38& 0.019&$\pm$0.015& 0.011&$\pm$0.015 &122.3\\
 \object{NGC4636}&150&32.59 &-5.76&$\pm$3.63&209.7&$\pm$4.57& 0.022&$\pm$0.016& 0.026&$\pm$0.016 &118.4\\
 \object{NGC4636}&150&36.58 &-3.04&$\pm$4.25&219.8&$\pm$5.31& 0.012&$\pm$0.018& 0.023&$\pm$0.018 &105.9\\
 \object{NGC4636}&150&41.04 & 4.61&$\pm$4.10&225.8&$\pm$4.61& 0.031&$\pm$0.016&-0.019&$\pm$0.016 &112.9\\ 
 \object{NGC4636}&150&46.20 & 1.35&$\pm$4.96&214.4&$\pm$5.22& 0.059&$\pm$0.021&-0.044&$\pm$0.021 &88.5\\ 
 \object{NGC4636}&150&52.07 & 7.14&$\pm$5.81&229.5&$\pm$6.10& 0.050&$\pm$0.023&-0.045&$\pm$0.023 &80.9\\
 \object{NGC4636}&150&58.63 &-3.40&$\pm$4.45&194.6&$\pm$3.06& 0.112&$\pm$0.021&-0.167&$\pm$0.021 &89.6\\
 \object{NGC4636}&150&66.15 &-10.48&$\pm$4.84&194.2&$\pm$4.73&0.107&$\pm$0.023&-0.069&$\pm$0.023 &82.2\\
 \object{NGC4636}&150&74.81 &-3.40&$\pm$6.26&204.3&$\pm$4.83& 0.121&$\pm$0.028&-0.139&$\pm$0.028 &66.9\\
 \object{NGC4636}&150&84.90 &-18.95&$\pm$5.63&204.1&$\pm$6.18&0.14&$\pm$0.025 &-0.029&$\pm$0.025 &74.3\\
 \object{NGC4636}&150&96.40 &-73.35&$\pm$4.50&190.8&$\pm$5.91&0.161&$\pm$0.027& 0.044&$\pm$0.027 &68.6\\
 \object{NGC4636}&150&109.70&-21.30&$\pm$7.34&191.2&$\pm$4.34&0.153&$\pm$0.035&-0.073&$\pm$0.035 &53.4\\
 \object{NGC4636}&150&126.20&-99.93&$\pm$7.01&147.3&$\pm$8.52&0.168&$\pm$0.043& 0.011&$\pm$0.043 &43.0\\
 \object{NGC4636}&150&137.18&-80.99&$\pm$14.87&138.3&$\pm$26.35&0.149&$\pm$0.098&0.133&$\pm$0.098 &39.1\\
% Minor axis 
 \object{NGC4636}&60& -5.29 &28.13 &$\pm$3.76&169.7&$\pm$4.53&-0.050&$\pm$0.020& 0.008&$\pm$0.020 &102.9\\
 \object{NGC4636}&60& -4.13 & 9.01 &$\pm$4.19&188.8&$\pm$4.57&-0.008&$\pm$0.020&-0.030&$\pm$0.020 &102.8\\
 \object{NGC4636}&60& -3.19 &15.98 &$\pm$4.05&185.4&$\pm$4.62&-0.035&$\pm$0.020&-0.014&$\pm$0.020 &104.3\\
 \object{NGC4636}&60& -2.25 &10.58 &$\pm$3.88&190.7&$\pm$4.45& 0.027&$\pm$0.019&-0.012&$\pm$0.019 &111.9\\
 \object{NGC4636}&60& -1.31 & 1.93 &$\pm$3.64&205.9&$\pm$3.89& 0.036&$\pm$0.016&-0.038&$\pm$0.016 &129.1\\
 \object{NGC4636}&60& -0.61 &-0.57 &$\pm$3.66&212.1&$\pm$3.88&-0.013&$\pm$0.016&-0.041&$\pm$0.016 &132.2\\
 \object{NGC4636}&60& -0.14 & 1.11 &$\pm$3.61&202.9&$\pm$3.77&-0.002&$\pm$0.016&-0.047&$\pm$0.016 &128.3\\
 \object{NGC4636}&60&  0.33 &-7.47 &$\pm$3.83&214.5&$\pm$4.18& 0.016&$\pm$0.016&-0.031&$\pm$0.016 &127.5\\
 \object{NGC4636}&60&  0.80 &-3.68 &$\pm$3.89&208.0&$\pm$4.74&-0.010&$\pm$0.017& 0.012&$\pm$0.017 &121.8\\
 \object{NGC4636}&60&  1.49 &-3.72 &$\pm$4.08&193.4&$\pm$4.39& 0.062&$\pm$0.019&-0.036&$\pm$0.019 &108.0\\
 \object{NGC4636}&60&  2.43 &-4.08 &$\pm$3.89&159.3&$\pm$2.40& 0.080&$\pm$0.022&-0.191&$\pm$0.022 &93.4\\
 \object{NGC4636}&60&  3.37 & 0.00 &$\pm$3.58&180.2&$\pm$4.12&-0.006&$\pm$0.018&-0.011&$\pm$0.018 &114.6\\
 \object{NGC4636}&60&  4.31 &13.04 &$\pm$4.68 &189.7&$\pm$5.49 &-0.020&$\pm$0.022 &-0.003&$\pm$0.022 &92.3\\
 \object{NGC4636}&60&  5.47 & 2.68 &$\pm$3.41 &150.9&$\pm$3.29 & 0.024&$\pm$0.021 &-0.074&$\pm$0.021 &100.8\\
 \object{NGC4636}&60&  7.10 &-4.08 &$\pm$3.49 &168.6&$\pm$4.00 &-0.049&$\pm$0.019 &-0.012&$\pm$0.019 &110.1\\
 \object{NGC4636}&60&  9.20 &14.11 &$\pm$3.74 &174.5&$\pm$4.30 &-0.049&$\pm$0.020 &-0.011&$\pm$0.020 &106.2\\
 \object{NGC4636}&60& 11.77 & 8.18 &$\pm$3.86 &168.2&$\pm$4.59 &-0.065&$\pm$0.021 & 0.003&$\pm$0.021 &99.3\\
 \object{NGC4636}&60& 15.03 &16.04 &$\pm$3.92 &136.0&$\pm$4.68 &-0.044&$\pm$0.026 & 0.005&$\pm$0.026 &90.3\\
 \object{NGC4636}&60& 19.46 & 5.39 &$\pm$3.56 &149.7&$\pm$4.44 &-0.070&$\pm$0.022 & 0.022&$\pm$0.022 &98.4\\
 \object{NGC4636}&60& 25.71 &17.65 &$\pm$4.58 &164.2&$\pm$4.58 &-0.126&$\pm$0.025 &-0.061&$\pm$0.025 &81.7\\
 \object{NGC4636}&60& 34.90 &19.90 &$\pm$6.52 &138.7&$\pm$6.76 &-0.027&$\pm$0.043 &-0.049&$\pm$0.043 &50.6\\
 \object{NGC4636}&60& 49.04 &56.54 &$\pm$10.23 &129.6&$\mp$6.04 &-0.138&$\pm$0.072 &-0.137&$\pm$0.072 &35.8\\
 \object{NGC4636}&60& 71.14 &45.24 &$\pm$13.98 &112.3&$\pm$8.26 &-0.104&$\pm$0.113 &-0.125&$\pm$0.113 &24.7\\
 \object{NGC4636}&60&101.22 &13.33 &$\pm$19.39 &111.5&$\pm$11.46 &-0.168&$\pm$0.158 &-0.201&$\pm$0.158 &16.6\\

\hline
\end{tabular}
\end{table*}

%LICK/IDS 
\begin{table*}
\begin{small}
\caption{The full table of measured Lick/IDS as a function of distance from the center 
(positive: east, negative: west) for the different position angles.} 
\begin{tabular}{lcrcccccc}
\hline
      Galaxy     &  PA   &     R   &Mgb$\pm$dMgb&Fe5015$\pm$dFe5015&Fe5270$\pm$dFe5270&Fe5335$\pm$dFe5335&Fe5406$\pm$dFe5406&H$\beta$$\pm$dH$\beta$\\
      Name       & (deg) &(\arcsec)&  (\AA)     &      (\AA)       &     (\AA)        &      (\AA)       &         (\AA)    &      (\AA)       \\
\hline
%                             Mgb     dMgb    Fe4     dFe4    Fe5     dFe5    Fe6     dFe6    Fe7     dFe7    Hbe     dHbe
\object{N4636}&150& -13.17&4.262$\pm$0.039&6.434$\pm$0.081&2.374$\pm$0.048&2.284$\pm$0.062&1.557$\pm$0.045&1.869$\pm$0.033\\
\object{N4636}&150& -11.77&4.341$\pm$0.026&4.515$\pm$0.055&2.478$\pm$0.031&2.375$\pm$0.040&1.577$\pm$0.030&1.880$\pm$0.022\\
\object{N4636}&150& -10.60&4.454$\pm$0.034&5.493$\pm$0.070&2.458$\pm$0.041&2.396$\pm$0.053&1.570$\pm$0.040&1.162$\pm$0.029\\
\object{N4636}&150& -9.66&4.445$\pm$0.030&5.497$\pm$0.062&2.562$\pm$0.037&2.436$\pm$0.047&1.606$\pm$0.035&1.702$\pm$0.026\\
\object{N4636}&150& -8.72&4.494$\pm$0.030&5.601$\pm$0.062&2.480$\pm$0.037&2.328$\pm$0.047&1.718$\pm$0.035&1.643$\pm$0.025\\
\object{N4636}&150& -7.78&4.597$\pm$0.030&5.520$\pm$0.062&2.514$\pm$0.037&2.529$\pm$0.047&1.655$\pm$0.036&1.494$\pm$0.025\\
\object{N4636}&150& -6.84&4.671$\pm$0.026&5.466$\pm$0.054&2.645$\pm$0.032&2.618$\pm$0.041&1.657$\pm$0.031&1.480$\pm$0.022\\
\object{N4636}&150& -5.90&4.752$\pm$0.028&5.532$\pm$0.057&2.632$\pm$0.034&2.517$\pm$0.043&1.766$\pm$0.033&1.425$\pm$0.024\\
\object{N4636}&150& -5.21&4.854$\pm$0.028&5.625$\pm$0.056&2.636$\pm$0.033&2.494$\pm$0.043&1.803$\pm$0.032&1.357$\pm$0.023\\
\object{N4636}&150& -4.74&4.811$\pm$0.032&5.573$\pm$0.065&2.712$\pm$0.038&2.565$\pm$0.050&1.705$\pm$0.037&1.393$\pm$0.027\\
\object{N4636}&150& -4.27&4.819$\pm$0.028&5.599$\pm$0.058&2.752$\pm$0.034&2.603$\pm$0.044&1.681$\pm$0.033&1.435$\pm$0.024\\
\object{N4636}&150& -3.80&4.898$\pm$0.027&5.644$\pm$0.056&2.804$\pm$0.033&2.730$\pm$0.043&1.769$\pm$0.032&1.368$\pm$0.023\\
\object{N4636}&150& -3.33&4.916$\pm$0.023&5.618$\pm$0.047&2.762$\pm$0.028&2.767$\pm$0.036&1.766$\pm$0.027&1.239$\pm$0.019\\
\object{N4636}&150& -2.86&4.957$\pm$0.026&5.702$\pm$0.053&2.712$\pm$0.031&2.726$\pm$0.041&1.742$\pm$0.031&1.208$\pm$0.022\\
\object{N4636}&150& -2.39&5.027$\pm$0.023&5.766$\pm$0.046&2.678$\pm$0.027&2.694$\pm$0.035&1.736$\pm$0.027&1.245$\pm$0.019\\
\object{N4636}&150& -1.92&5.099$\pm$0.021&5.708$\pm$0.043&2.795$\pm$0.025&2.769$\pm$0.033&1.770$\pm$0.025&1.221$\pm$0.017\\
\object{N4636}&150& -1.45&5.111$\pm$0.020&5.623$\pm$0.042&2.709$\pm$0.025&2.800$\pm$0.032&1.764$\pm$0.024&1.220$\pm$0.017\\
\object{N4636}&150& -0.98&5.125$\pm$0.021&5.640$\pm$0.043&2.680$\pm$0.025&2.811$\pm$0.033&1.780$\pm$0.025&1.273$\pm$0.018\\
\object{N4636}&150& -0.51&5.098$\pm$0.021&5.755$\pm$0.044&2.706$\pm$0.026&2.818$\pm$0.034&1.813$\pm$0.026&1.351$\pm$0.018\\
\object{N4636}&150& -0.04&5.083$\pm$0.022&6.045$\pm$0.044&2.759$\pm$0.026&2.841$\pm$0.035&1.811$\pm$0.026&1.400$\pm$0.018\\
\object{N4636}&150&  0.44&5.008$\pm$0.020&6.095$\pm$0.042&2.774$\pm$0.025&2.811$\pm$0.032&1.816$\pm$0.024&1.449$\pm$0.017\\
\object{N4636}&150&  0.91&4.957$\pm$0.020&6.140$\pm$0.042&2.794$\pm$0.025&2.827$\pm$0.032&1.795$\pm$0.024&1.498$\pm$0.017\\
\object{N4636}&150&  1.38&4.950$\pm$0.022&6.113$\pm$0.045&2.803$\pm$0.026&2.792$\pm$0.034&1.762$\pm$0.026&1.553$\pm$0.018\\
\object{N4636}&150&  1.85&4.964$\pm$0.022&5.986$\pm$0.046&2.759$\pm$0.027&2.818$\pm$0.035&1.755$\pm$0.026&1.571$\pm$0.018\\
\object{N4636}&150&  2.32&4.958$\pm$0.025&5.866$\pm$0.053&2.719$\pm$0.031&2.796$\pm$0.040&1.756$\pm$0.030&1.528$\pm$0.022\\
\object{N4636}&150&  2.79&4.950$\pm$0.025&5.835$\pm$0.051&2.712$\pm$0.030&2.750$\pm$0.039&1.816$\pm$0.029&1.453$\pm$0.021\\
\object{N4636}&150&  3.26&4.919$\pm$0.029&5.722$\pm$0.059&2.671$\pm$0.035&2.697$\pm$0.045&1.868$\pm$0.034&1.488$\pm$0.024\\
\object{N4636}&150&  3.73&4.876$\pm$0.026&5.716$\pm$0.052&2.634$\pm$0.031&2.756$\pm$0.040&1.755$\pm$0.030&1.592$\pm$0.022\\
\object{N4636}&150&  4.20&4.814$\pm$0.026&5.797$\pm$0.053&2.656$\pm$0.031&2.668$\pm$0.040&1.685$\pm$0.030&1.683$\pm$0.022\\
\object{N4636}&150&  4.67&4.741$\pm$0.030&5.791$\pm$0.061&2.570$\pm$0.036&2.555$\pm$0.046&1.654$\pm$0.035&1.685$\pm$0.025\\
\object{N4636}&150&  5.14&4.669$\pm$0.031&5.849$\pm$0.065&2.674$\pm$0.038&2.668$\pm$0.049&1.688$\pm$0.037&1.759$\pm$0.026\\
\object{N4636}&150&  5.83&4.650$\pm$0.023&5.973$\pm$0.047&2.688$\pm$0.028&2.567$\pm$0.036&1.661$\pm$0.027&1.957$\pm$0.019\\
\object{N4636}&150&  6.77&4.682$\pm$0.024&5.972$\pm$0.048&2.686$\pm$0.029&2.636$\pm$0.037&1.633$\pm$0.028&1.789$\pm$0.020\\
\object{N4636}&150&  7.71&4.580$\pm$0.027&5.867$\pm$0.056&2.585$\pm$0.033&2.572$\pm$0.042&1.644$\pm$0.032&1.479$\pm$0.023\\
\object{N4636}&150&  8.65&4.478$\pm$0.032&5.693$\pm$0.064&2.580$\pm$0.038&2.394$\pm$0.049&1.582$\pm$0.037&1.464$\pm$0.027\\
\object{N4635}&150&  9.59&4.490$\pm$0.028&5.702$\pm$0.057&2.600$\pm$0.034&2.451$\pm$0.044&1.666$\pm$0.033&1.494$\pm$0.023\\
\object{N4636}&150& 10.76&4.279$\pm$0.030&5.600$\pm$0.061&2.409$\pm$0.036&2.239$\pm$0.047&1.580$\pm$0.035&1.736$\pm$0.025\\
\object{N4636}&150& 12.17&4.280$\pm$0.033&5.802$\pm$0.069&2.418$\pm$0.041&2.235$\pm$0.052&1.561$\pm$0.039&1.900$\pm$0.028\\
\object{N4636}&150& 13.58&4.208$\pm$0.036&5.506$\pm$0.072&2.560$\pm$0.043&2.355$\pm$0.056&1.643$\pm$0.042&1.844$\pm$0.029\\
\object{N4636}&150& 15.22&4.137$\pm$0.033&5.612$\pm$0.066&2.427$\pm$0.039&2.300$\pm$0.051&1.573$\pm$0.038&1.829$\pm$0.027\\
\object{N4636}&150& 17.10&4.105$\pm$0.040&5.803$\pm$0.083&2.376$\pm$0.049&2.386$\pm$0.063&1.485$\pm$0.047&1.808$\pm$0.033\\
\object{N4636}&150& 18.98&4.056$\pm$0.044&5.549$\pm$0.088&2.335$\pm$0.052&2.398$\pm$0.068&1.712$\pm$0.051&1.857$\pm$0.036\\
\object{N4636}&150& 21.09&3.880$\pm$0.037&5.920$\pm$0.075&2.403$\pm$0.044&2.227$\pm$0.057&1.491$\pm$0.043&1.887$\pm$0.031\\
\object{N4636}&150& 23.44&3.915$\pm$0.039&5.522$\pm$0.080&2.263$\pm$0.048&2.190$\pm$0.062&1.570$\pm$0.046&1.930$\pm$0.033\\
\hline
\end{tabular}
\label{tab_lick}
\end{small}
\end{table*}

\addtocounter{table}{-1}
\begin{table*}
\begin{small}
\caption{Continued} 
\begin{tabular}{lcrcccccc}
\hline
      Galaxy     &  PA   &     R   &Mgb$\pm$dMgb&Fe5015$\pm$dFe5015&Fe5270$\pm$dFe5270&Fe5335$\pm$dFe5335&Fe5406$\pm$dFe5406&H$\beta$$\pm$dH$\beta$\\
      Name       & (deg) &(\arcsec)&  (\AA)     &      (\AA)       &     (\AA)        &      (\AA)       &         (\AA)    &      (\AA)       \\
\hline

\object{N4636}&150& 26.02&3.884$\pm$0.046&5.426$\pm$0.093&2.349$\pm$0.055&2.199$\pm$0.072&1.503$\pm$0.054&1.774$\pm$0.038\\
\object{N4636}&150& 29.07&3.870$\pm$0.040&5.451$\pm$0.081&2.258$\pm$0.049&2.142$\pm$0.062&1.492$\pm$0.047&1.918$\pm$0.033\\
\object{N4636}&150& 32.59&3.710$\pm$0.041&5.381$\pm$0.085&2.398$\pm$0.049&2.058$\pm$0.064&1.401$\pm$0.048&1.987$\pm$0.034\\
\object{N4636}&150& 36.58&3.620$\pm$0.046&4.876$\pm$0.096&2.439$\pm$0.056&2.056$\pm$0.073&1.420$\pm$0.055&2.002$\pm$0.038\\
\object{N4636}&150& 41.04&3.529$\pm$0.044&5.369$\pm$0.088&2.297$\pm$0.052&2.017$\pm$0.070&1.478$\pm$0.052&1.772$\pm$0.036\\
\object{N4636}&150& 46.20&3.705$\pm$0.055&4.820$\pm$0.112&2.319$\pm$0.066&1.908$\pm$0.087&1.389$\pm$0.065&1.554$\pm$0.045\\
\object{N4636}&150& 52.07&3.483$\pm$0.061&5.256$\pm$0.124&2.334$\pm$0.073&2.086$\pm$0.098&1.360$\pm$0.073&1.902$\pm$0.050\\
\object{N4636}&150& 58.63&3.330$\pm$0.054&3.758$\pm$0.112&2.123$\pm$0.065&2.021$\pm$0.082&1.268$\pm$0.062&1.885$\pm$0.045\\
\object{N4636}&150& 66.15&3.156$\pm$0.059&4.549$\pm$0.119&2.281$\pm$0.071&1.724$\pm$0.090&1.501$\pm$0.067&1.746$\pm$0.049\\
\object{N4636}&150& 74.81&2.927$\pm$0.073&4.546$\pm$0.150&2.274$\pm$0.087&1.910$\pm$0.112&1.419$\pm$0.084&1.698$\pm$0.060\\
\object{N4636}&150& 84.90&3.210$\pm$0.066&4.632$\pm$0.133&2.518$\pm$0.079&2.067$\pm$0.102&1.435$\pm$0.075&2.001$\pm$0.054\\
\object{N4636}&150& 96.40&2.813$\pm$0.071&4.104$\pm$0.143&2.383$\pm$0.085&1.877$\pm$0.107&1.331$\pm$0.080&2.043$\pm$0.058\\
\object{N4636}&150& 109.70&2.452$\pm$0.092&4.197$\pm$0.183&2.526$\pm$0.109&1.631$\pm$0.140&1.487$\pm$0.103&2.127$\pm$0.075\\
\object{N4636}&150& 126.20&2.995$\pm$0.110&3.588$\pm$0.221&2.544$\pm$0.132&1.493$\pm$0.163&1.265$\pm$0.120&2.040$\pm$0.093\\
\object{N4636}&150& 137.18&3.403$\pm$0.243&3.139$\pm$0.494&2.376$\pm$0.290&2.315$\pm$0.357&1.559$\pm$0.266&2.171$\pm$0.209\\
\object{N4636}&60&-5.290&4.474$\pm$0.045&4.803$\pm$0.094&2.761$\pm$0.055&2.661$\pm$0.068&1.681$\pm$0.051&2.792$\pm$0.039\\
\object{N4636}&60&-4.13&4.720$\pm$0.046&4.676$\pm$0.096&2.962$\pm$0.056&2.600$\pm$0.072&1.754$\pm$0.053&1.195$\pm$0.040\\
\object{N4636}&60&-3.19&4.884$\pm$0.045&4.773$\pm$0.094&2.936$\pm$0.055&2.556$\pm$0.070&1.774$\pm$0.052&1.117$\pm$0.039\\
\object{N4636}&60&-2.25&5.009$\pm$0.043&4.928$\pm$0.088&2.952$\pm$0.052&2.566$\pm$0.066&1.892$\pm$0.049&1.019$\pm$0.037\\
\object{N4636}&60&-1.31&5.019$\pm$0.037&5.138$\pm$0.079&3.068$\pm$0.045&2.719$\pm$0.058&1.941$\pm$0.043&1.028$\pm$0.032\\
\object{N4636}&60&-0.61&5.070$\pm$0.036&5.329$\pm$0.077&3.101$\pm$0.044&2.835$\pm$0.057&1.886$\pm$0.043&1.053$\pm$0.031\\
\object{N4636}&60&-0.14&5.090$\pm$0.037&5.196$\pm$0.079&3.128$\pm$0.045&2.784$\pm$0.058&1.870$\pm$0.044&0.984$\pm$0.032\\
\object{N4636}&60& 0.33&5.160$\pm$0.038&5.202$\pm$0.079&3.075$\pm$0.046&2.908$\pm$0.060&1.953$\pm$0.045&0.941$\pm$0.032\\
\object{N4636}&60& 0.80&5.028$\pm$0.039&5.259$\pm$0.082&2.882$\pm$0.048&2.791$\pm$0.062&1.842$\pm$0.046&0.969$\pm$0.034\\
\object{N4636}&60& 1.49&4.986$\pm$0.044&4.869$\pm$0.091&3.058$\pm$0.053&2.561$\pm$0.068&1.817$\pm$0.051&1.017$\pm$0.038\\
\object{N4636}&60& 2.43&4.869$\pm$0.049&4.955$\pm$0.102&2.790$\pm$0.060&2.355$\pm$0.074&1.784$\pm$0.056&0.953$\pm$0.044\\
\object{N4636}&60& 3.37&4.922$\pm$0.041&5.081$\pm$0.086&2.791$\pm$0.050&2.534$\pm$0.063&1.752$\pm$0.047&0.717$\pm$0.036\\
\object{N4636}&60& 4.31&4.966$\pm$0.051&4.781$\pm$0.106&2.997$\pm$0.063&2.658$\pm$0.080&1.803$\pm$0.059&0.719$\pm$0.044\\
\object{N4636}&60& 5.47&4.675$\pm$0.046&4.430$\pm$0.094&2.798$\pm$0.055&2.433$\pm$0.068&1.684$\pm$0.051&0.946$\pm$0.040\\
\object{N4636}&60& 7.10&4.527$\pm$0.042&5.267$\pm$0.087&2.620$\pm$0.051&2.266$\pm$0.064&1.726$\pm$0.048&1.237$\pm$0.037\\
\object{N4636}&60& 9.20&4.507$\pm$0.044&4.910$\pm$0.091&2.631$\pm$0.054&2.211$\pm$0.067&1.632$\pm$0.050&1.255$\pm$0.038\\
\object{N4636}&60&11.77&4.309$\pm$0.047&4.841$\pm$0.097&2.349$\pm$0.057&2.257$\pm$0.072&1.420$\pm$0.053&1.716$\pm$0.041\\
\object{N4636}&60&15.03&4.223$\pm$0.051&4.611$\pm$0.104&2.387$\pm$0.062&2.343$\pm$0.075&1.286$\pm$0.056&1.640$\pm$0.045\\
\object{N4636}&60&19.46&4.075$\pm$0.047&4.622$\pm$0.097&2.245$\pm$0.057&2.076$\pm$0.070&1.461$\pm$0.052&1.759$\pm$0.041\\
\object{N4636}&60&25.71&3.768$\pm$0.057&4.601$\pm$0.118&2.205$\pm$0.070&2.221$\pm$0.086&1.358$\pm$0.064&1.443$\pm$0.050\\
\object{N4636}&60&34.90&3.186$\pm$0.092&4.497$\pm$0.188&1.677$\pm$0.112&2.255$\pm$0.135&1.191$\pm$0.100&1.664$\pm$0.080\\
\object{N4636}&60&49.04&3.164$\pm$0.130&3.836$\pm$0.265&1.775$\pm$0.157&2.467$\pm$0.186&1.159$\pm$0.140&1.949$\pm$0.113\\
\object{N4636}&60&71.14&2.607$\pm$0.188&3.477$\pm$0.381&0.940$\pm$0.223&2.139$\pm$0.266&1.051$\pm$0.201&1.427$\pm$0.165\\
\object{N4636}&60&101.22&2.575$\pm$0.279&3.659$\pm$0.568&0.478$\pm$0.336&2.835$\pm$0.395&1.071$\pm$0.296&3.797$\pm$0.239\\
\hline
\end{tabular}
\end{small}
\end{table*}

\end{document}